\documentclass[twocolumn,twoside,slac_two]{revtex4}
\usepackage{graphicx}
\usepackage{fancyhdr}
\begin{document}

\title{Results from the observation of extragalactic objects with the MAGIC telescope}

%

\author{D.~Kranich on behalf of the MAGIC collaboration}
\affiliation{ETH Zurich, CH-8093 Switzerland}

\begin{abstract}
In the last couple of years the Magic air Cherenkov telescope has
made significant contributions to very high energy $\gamma$-ray astronomy.
These include the detection of the galactic binary system LSI +61 303 and the
observation of pulsed emission from the Crab pulsar. Extragalactic objects
like the famous FSRQ 3C 279 and the LBL S5 0716+714 have both been detected
during optical high states, and the radio galaxy M87 could be observed during
an unexpected strong $\gamma$-ray outburst. Given its low energy trigger
threshold (~50 GeV) and fast repositioning time of less than 30s the Magic
air Cherenkov telescope is particularly well suited for the observation of
fast transient objects like AGN or GRBs. So far no GRB could be detected with
Magic, however. In this paper we present selected highlights from recent
MAGIC observations of extragalactic objects.

\end{abstract}

\maketitle

\thispagestyle{fancy}


\section{The MAGIC telescope}

\begin{figure}
\includegraphics[width=7cm]{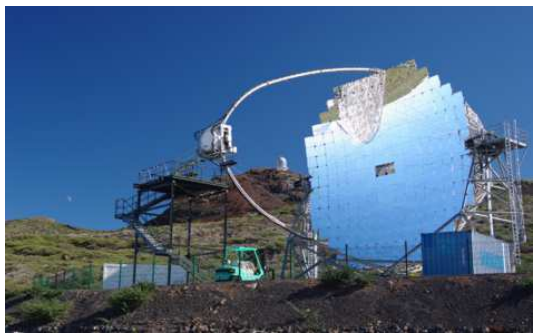}
\caption{The MAGIC telescope}
\label{MagicTelescope}
\end{figure}

The MAGIC Cherenkov telescope is located on the Canary Island La
Palma. The telescope has a large parabolic Mirror of { $234
 \mathrm{m}^2$}, a { $3.5^\circ$ FoV} camera with 577 enhanced QE
PMTs and a { 2 GS/s FADC} system. Its design was optimized for a
low 50 GeV trigger threshold and a fast repositioning time of $300^\circ$
in 30 s. MAGIC has an $\sim 30 \%$ { enhanced duty cycle} as observations
are also carried out during twilight or the presence of moonlight.
Further details on the telescope and its performance can be found in
\cite{baixeras}, \cite{cortina05} and \cite{Albert08a}.
A second telescope, MAGIC-II, which will significantly enhance the detector
performance is about to be fully commissioned \cite{moralejo}.

\section{Gamma Ray Bursts}

MAGIC started to follow-up GCN alerts in the beginning of 2005 and has
observed more than 50 GRBs since then. Out of 23 GRBs observed during
the last two years, 7 GRBs with accurate pointing position and weather
condition were thoroughly analyzed. No detection, but upper limits for
different energy regions could be made (Tab. \ref{GRB-Results}). Follow
up observations on accessible GRBs started in between 1 and 3 minutes
after the burst onset.

\begin{table}[t]
\begin{center}
\caption{ Upper limits for selected GRBs at different energy bins, given
in units of $10^{-10} \, \mathrm{erg} \, \mathrm{cm}^{-2} \,
\mathrm{s}^{-1}$.}
\begin{tabular}{|c|c|c|c|c|}
\hline
observation time &  \multicolumn{4}{c|}{Energy bin [GeV]} \\  \cline{2-5}
\raisebox{1.5ex}[0pt]{after $T_0$ in [s]}  	& 80-125 & 125-175 & 175-300 & 300-1000 \\ \hline

\multicolumn{5}{|c|}{GRB080315} \\ 
 160 $\rightarrow$ 1716		& --		& 0.16 	& 0.18 	& 0.05 \\
 1761 $\rightarrow$ 5061 	& --		& -- 		& 0.05 	& 0.05 \\ \hline

\multicolumn{5}{|c|}{GRB080319A} \\
 259 $\rightarrow$ 1736		& --		& 0.57 	& 0.11 	& 0.06 \\ \hline

\multicolumn{5}{|c|}{GRB080330} \\
 91 $\rightarrow$ 974		& --		& -- 		& 0.76 	& 0.49 \\ 
 974 $\rightarrow$ 1754		& --		& --	 	& --	 	& 0.33 \\ \hline

\multicolumn{5}{|c|}{GRB080430} \\
 4753 $\rightarrow$ 11011	& 0.55	& 0.77 	& 0.30 	& 0.05 \\ 
 16912 $\rightarrow$ 17968	& 7.76	& 6.59	& 2.45	& 0.92 \\ \hline

\multicolumn{5}{|c|}{GRB080603B} \\
 5578 $\rightarrow$ 7317	& --		& 0.16 	& 0.06 	& 0.01 \\ 
 7497 $\rightarrow$ 12357	& --		& 0.05	& 0.12	& 0.03 \\ \hline

\multicolumn{5}{|c|}{GRB090102} \\
 1161 $\rightarrow$ 5181	& 2.30	& 1.61 	& 0.34 	& 0.04 \\ 
 5181 $\rightarrow$ 9381	& 1.12	& 0.50	& 0.20	& 0.07 \\
 9861 $\rightarrow$ 11241	& --		& 1.02	& 0.30	& 0.20 \\
 11241 $\rightarrow$ 12621	& --		& 2.40	& 0.49	& 0.25 \\
 12621 $\rightarrow$ 14841	& --		& --		& 0.40	& 0.11 \\ \hline

\multicolumn{5}{|c|}{GRB090113} \\
 4603 $\rightarrow$ 10063	& 0.54	& 0.99 	& 0.10 	& 0.08 \\ 
 10183 $\rightarrow$ 13363	& --		& 0.93	& 0.2	3	& 0.08 \\ \hline
\hline
\end{tabular}
\label{GRB-Results}
\end{center}
\end{table}

\section{FSRQ 3C 279}

\begin{figure}
\includegraphics[width=7cm]{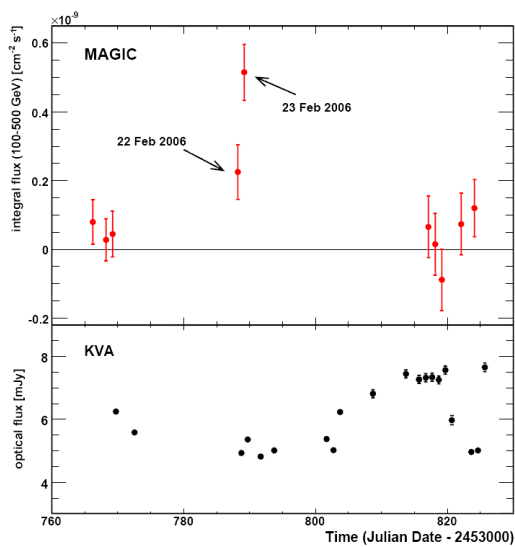}
\caption{VHE $\gamma$-ray (top) and optical R-band lightcurve (bottom) from
3C 279, taken in spring 2006.}
\label{3c279-lc}
\end{figure}

The FSRQ 3C 279 is one of the brightest objects at all wavelengths and was
the first blazar to be discovered at $\gamma$-rays \cite{hartman}. It is
also the first FSRQ found at VHE $\gamma$-rays \cite{Aliu2008} and, given
its redshift of $z=0.536$, the most distant VHE $\gamma$-ray emitter
detected so far. Due to its large redshift 3C 279 is an important object
to study the EBL photon field. The Magic observations which lead to the
detection at $5.8\, \sigma$ were performed in spring 2006. The signal
basically came from a single flare on February 23rd (Fig. \ref{3c279-lc}).
The optical and VHE $\gamma$-ray flux do not show correlated short-term
behavior during the campaign. As the overall optical flux was about a
factor 2 above the long-term baseline of ~3 mJy, this data do also support
some weak optical - VHE $\gamma$-ray correlation as found in other objects.

New observations after another optical outburst in January 2007 also led
to a $5.2\, \sigma$ detection \cite{karsten}.

\begin{figure}
\includegraphics[width=7cm]{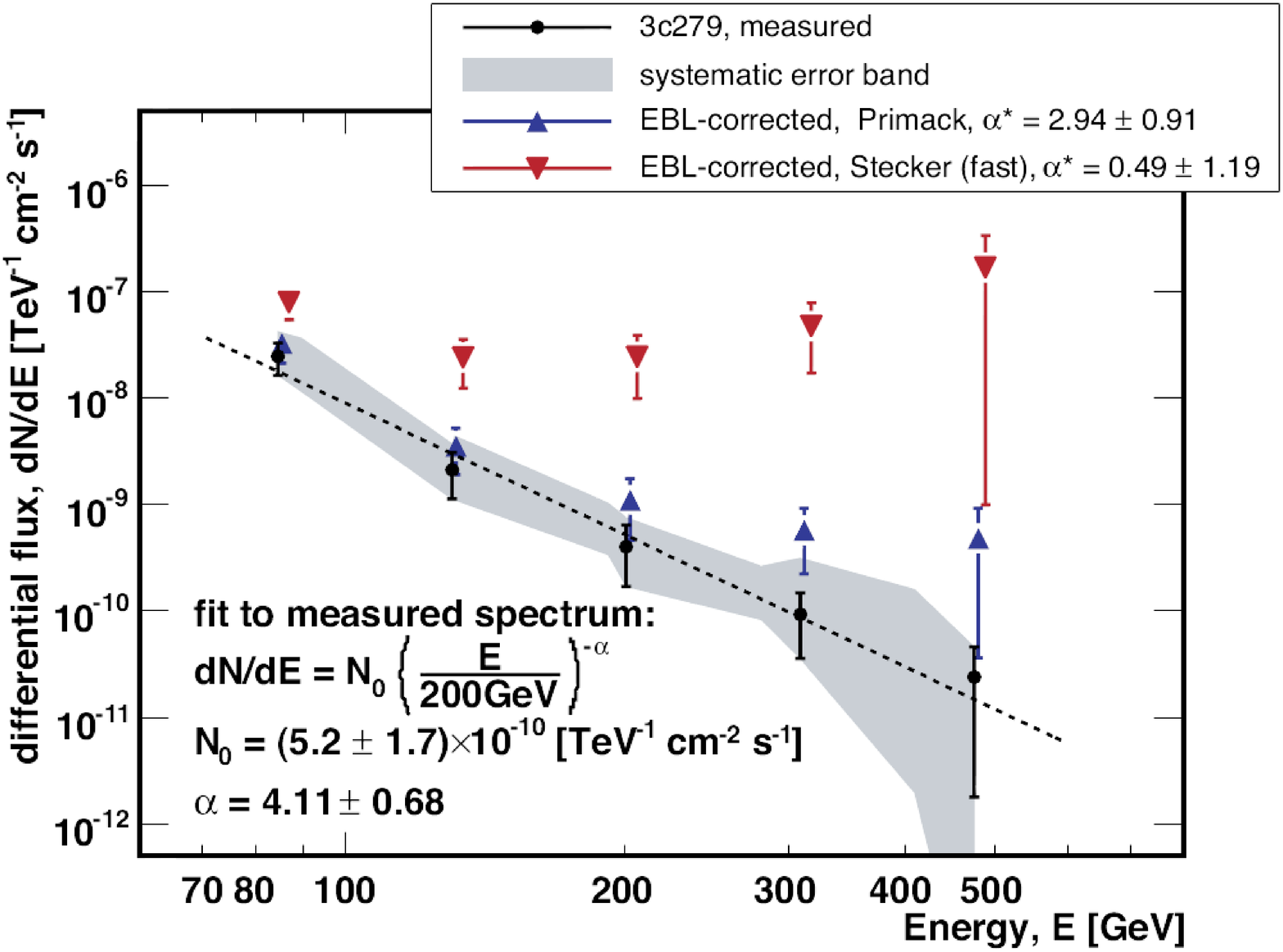}
\caption{MAGIC 3C 279 VHE $\gamma$-ray spectrum as obtained from the 2006 data. The red
and blue points refer to the de-aborbed spectra using two different EBL models.}
\label{3c279-spectrum}
\end{figure}

\begin{figure}
\includegraphics[width=7cm]{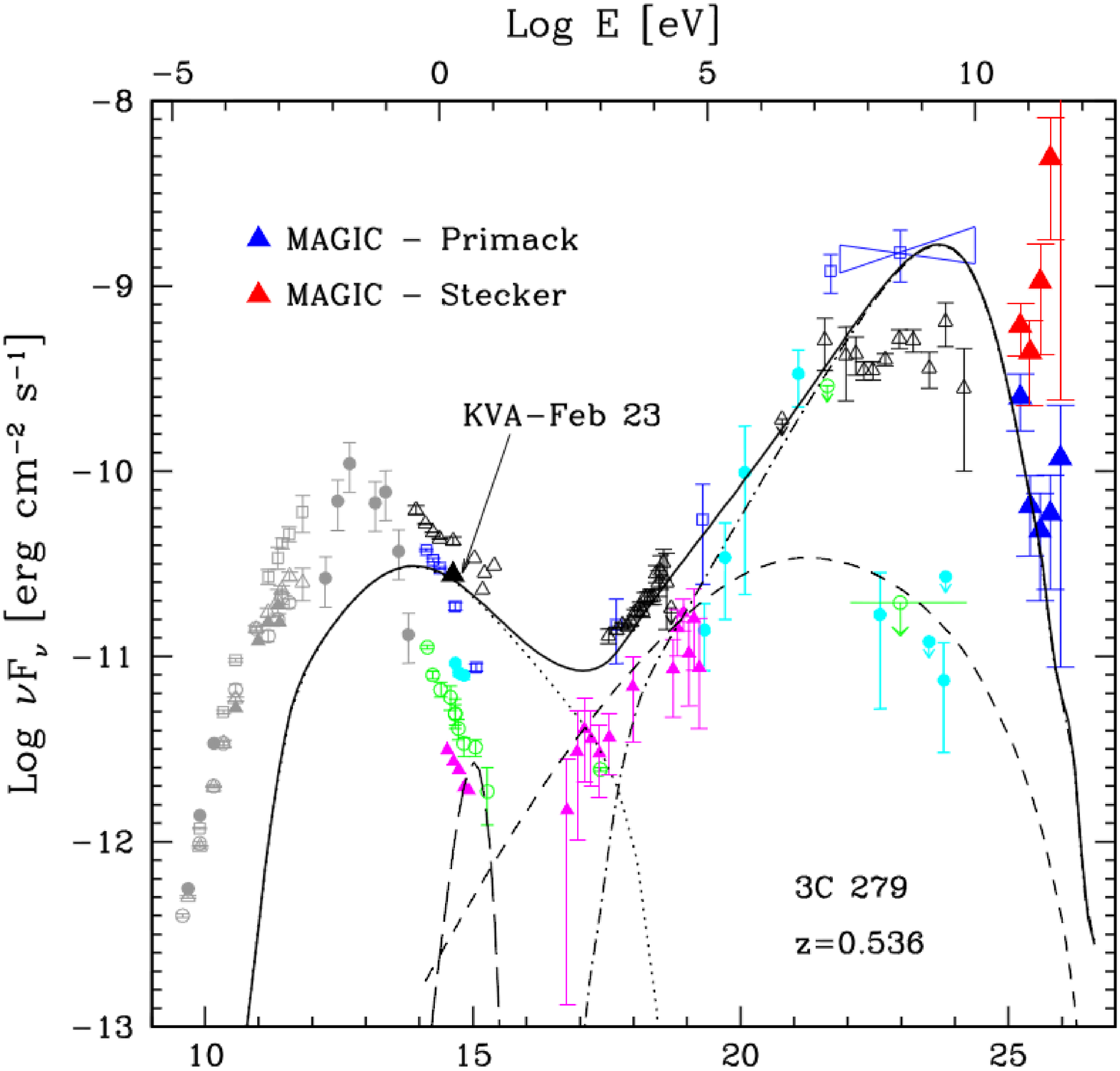}
\caption{Historical 3C 279 broad band SED (bottom) for the years 1991 to 2003
\cite{ballo, collmar} together with the 2006 MAGIC data points. The red and blue
MAGIC points refer to the de-aborbed spectra using two different EBL models.}
\label{3c279-ssc}
\end{figure}

The MAGIC energy spectrum is well described by a pure power law with photon
Index $\Gamma = 4.11 \pm 0.68$ (Fig. \ref{3c279-spectrum}). The low level EBL
corrected spectrum can be fitted by a one-zone SSC+EC model \cite{Maraschi2003}. 
Taking into account quasi-simultaneous optical and X-ray data, however, suggests
a multi-zone emission region (Fig. \ref{3c279-ssc}). Hadronic models also seem to
describe the data well \cite{Boettcher2008}.

\section{BL Lac S5 0716+714}

\begin{figure}
\includegraphics[width=7cm]{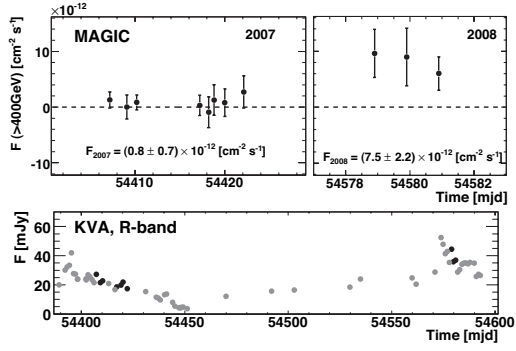}
\caption{The S5 0716+714 lightcurve in VHE $\gamma$-rays (upper panel) and
optical (lower panel) as measured from November 2007 until April 2008. The
optical data which was taken simultaneously with MAGIC are shown in black.}
\label{0716-lc}
\end{figure}

The low peaked BL Lac S5 0716+714 was observed with MAGIC for about 13h
between November 2007 and April 2008. Both observations were triggered
by strong optical flares seen by KVA. In November 2007 MAGIC follow up
observations began 11 days after the trigger due to moon and bad weather
constrains. The optical flux had already decreased significantly by then
and no signal could be found. The April 2008 observations started 5 days
after the trigger when the optical flux was still high (Fig. \ref{0716-lc}).
The significance of the combined data was $5.8 \, \sigma$ \cite{Anderhub2009}
mainly coming from the 2008 data set (giving $6.9 \, \sigma$). In 2008, the
integral flux above $400 \, $GeV was $F = 7.5 \pm 2.2\, 10^{-12}\,$ph/cm$^2$/s
which corresponds to about 9\% of the Crab Nebula flux. It should be noted
that S5 0716+714 was also in a historical high state in X-rays \cite{giommi}
in April 2008.

\begin{figure}
\includegraphics[width=7cm]{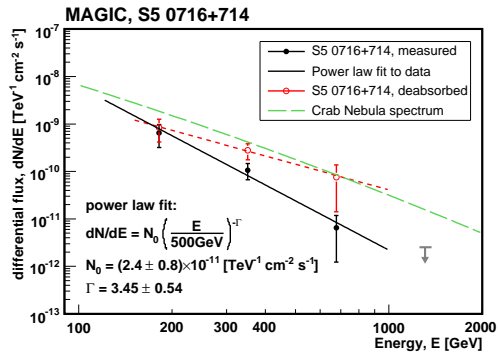}
\caption{S5 0716+714 energy spectrum as obtained from the 2008 data set. The
de-absorbed spectrum was derived using the EBL model of \cite{franceschini}
and assuming $z=0.26$}
\label{0716-spectrum}
\end{figure}

S5 0716+714 is the third low peaked VHE Blazar after BL Lac (MAGIC) and
W Comae (Veritas). The redshift was determined from the host galaxy as
$z=0.31\pm0.08$ \cite{Nilsson2008} which makes S5 0716+714 the 2nd farthest
VHE object detected so far. Due to the large redshift, the energy spectrum is
rather steep with a photon index $\Gamma = 3.45 \pm 0.54$ (Fig. \ref{0716-spectrum}).

\begin{figure}
\includegraphics[width=7cm]{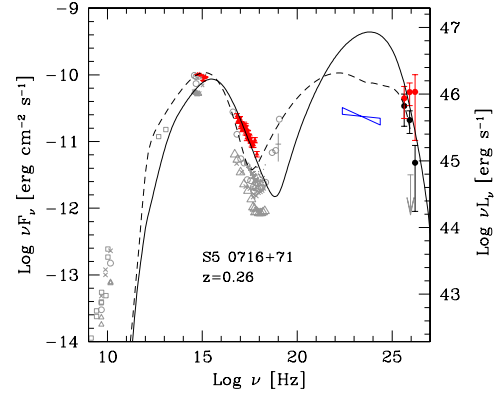}
\caption{S5 0716+714 SED for the 2008 data sample (red) and some historical
data (gray). The gray arrow corresponds to the MAGIC upper limit derived from the
November 2007 data sample. The result from a one-zone SSC model fit is shown as
solid line. The dashed line corresponds to a spine-layer model fit as described in
\cite{tavecchio2009}.}
\label{0716-ssc}
\end{figure}

The broadband SED for the April 2008 data set is shown in Fig. \ref{0716-ssc}.
A simple one-zone SSC model (solid line) well describes the broad band SED, but
predicts huge GeV fluxes. A structured jet model \cite{tavecchio2009}, shown
as dashed line, would not require such large GeV fluxes and, in addition, would
result in a photon index at GeV energies which is consistent with the one measured
by Fermi (blue bow tie in Fig. \ref{0716-ssc}) during a presumably different intensity
state.

\section{Mrk 501 Multi-Wavelength campaign}

\begin{figure}
\includegraphics[width=7cm]{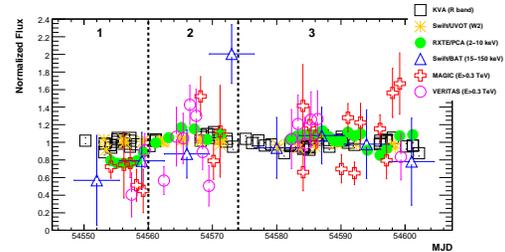}
\caption{Normalized Mrk 501 light curves as obtained for a selection of the instruments
taking part in the campaign 2008. The vertical bars denote $1\ \sigma$  statistical
uncertainties, and the horizontal bars the integration time of the observation.}
\label{501-lc}
\end{figure}

\begin{figure}
\includegraphics[width=7cm]{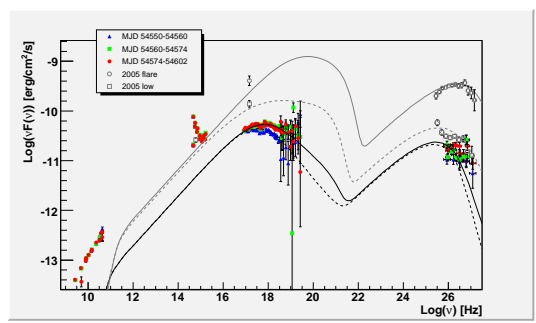}
\caption{Broadband SED for Mrk 501 as obtained during the 2008 MWL campaign in comparison
to low and high states from 2005. The results from a SSC model fit to the low state 2008
data (dot-dashed cuve), the high state 2008 data (heavy-dashed curve), the 2005 low state
(light-dashed curve) and the 2005 high state (solid line curve) are also shown.}
\label{501-sed}
\end{figure}

Mrk 501 was observed between March and May 2008 as part of an extensive multiwavelength
observation campaign covering radio (Effelsberg, IRAM, Medicina, Mets\"ahovi, Noto,
RATAN-600, VLBA), optical (KVA), UV (Swift/UVOT), X-ray (RXTE/PCA, Swift/XRT and Swift/BAT)
and $\gamma$-ray (MAGIC, Whipple, VERITAS) energies. The duration as well as the energy
coverage of this particular Mrk 501 campaign are rather unique. Mrk 501 was found in a
low state of activity at about 20\% the Crab Nebula VHE flux. Nevertheless, significant
flux variations could be observed. In particular at X-rays but also, less pronounced, in
$\gamma$-rays (Fig. \ref{501-lc}). Overall Mrk 501 showed an increased variability when
going from radio to $\gamma$-ray energies.\\
Significant correlations between different energy bands have been found for the pairs
RXTE/PCA - Swift/XRT (DCF: $0.87\pm0.28$) and also, less significant, RXTE/PCA (or Swift/XRT)
with VHE $\gamma$-rays (DCF: $0.5\pm0.19$). No hints for possible time-lags have been found.
The correlation analysis was, however, strongly affected by the modest flux variability
and / or the large flux errors.
The broadband spectral energy distribution during the two different emission states of
the campaign is well described by a homogeneous one-zone synchrotron self-Compton model
\cite{Tavecchio2001} as shown in Fig. \ref{501-sed}. The high emission state was
satisfactorily modeled by increasing the amount of high energy electrons with respect to
the low emission state. This parameterization is consistent with the energy-dependent
variability trend observed during the campaign. The model parameter values for the
different emission states are given in table \ref{501-ssc}. 

\begin{table}[t]
\begin{center}
\caption{The SSC model parameters used to describe the broadband SED for different
flux states of the Mrk 501 MWL campaign in 2005 and 2008.}
\begin{tabular}{|c|c|c|c|c|}
\hline
    SSC parameters&  2008 high & 2008 low& 2005 high& 2005 low\\
    \hline
    &&&&\\
    $\gamma_{\mathrm{break}}$ & $2.6 \cdot 10^5$  &
    $2.2 \cdot 10^5$ & $1.0 \cdot 10^6$  & $1.0 \cdot 10^5$\\ 
    $e^-$ slope n$_1$ & 2.0 & 2.0 & 2.0 & 2.0\\
    $e^-$ slope n$_2$ & 3.9 & 4.2 & 3.9 & 3.2\\
    mag. field B [G] & 0.19 & 0.19 & 0.23 & 0.31 \\
    density K [cm$^{-3}]$ & $1.8 \cdot 10^4$ & $1.8 \cdot 10^4$ & $7.5 \cdot
    10^4$ & $4.3 \cdot 10^4$ \\
    Radius R [cm] & $3 \cdot 10^{15}$ & $3 \cdot 10^{15}$ & $1 \cdot 10^{15}$
    & $1 \cdot 10^{15}$ \\
    Doppler factor $\delta$ & 12 & 12 & 25 & 25\\
    \hline
\hline
\end{tabular}
\label{501-ssc}
\end{center}
\end{table}

\section{Radio Galaxy M 87}

M 87 is a nearby Radio Galaxy located in the center of the Virgo cluster and was the
first Radio Galaxy to be detected at VHE $\gamma$-rays \cite{Aharonian}. In 2008 a
joint campaign involving all three next generation Cherenkov telescopes (H.E.S.S., MAGIC
and VERITAS) observed M 87 for in total 120h from January until May. In February, MAGIC
detected a strongly increased emission from M 87 and triggered MWL observations with
radio, X-ray and extended VHE coverage \cite{Acciari2009}. The data show evidence for a
correlated radio and VHE $\gamma$-ray emission and allowed to constrain the origin of the
VHE $\gamma$-ray emission to the immediate vincinity of the central black hole. There
is a dedicated presentation \cite{wagner} on this campaign.

\begin{figure}
\includegraphics[width=7cm]{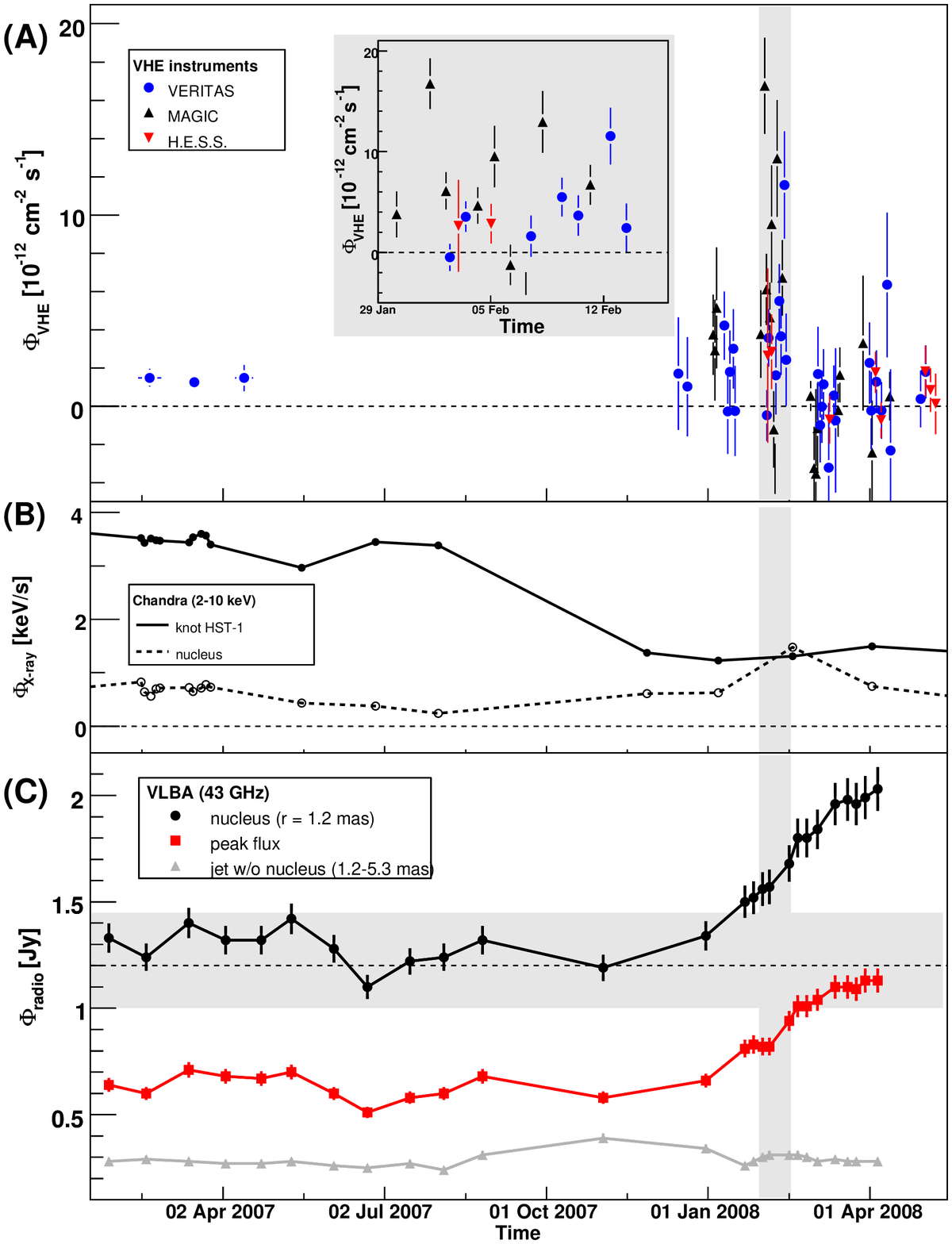}
\caption{M87 light curves from 2007 to 2008 at VHE $\gamma$-rays (top,
H.E.S.S., MAGIC and VERITAS), X-rays (middle, Chandra) and radio (bottom,
43 GHz, VLBA). The inlay in the top panel shows a zoom of the flare in
February 2008.}
\label{M87-lc}
\end{figure}

\bigskip 
\begin{acknowledgments}
We thank the Instituto de Astrofisica de Canarias for the excellent
working conditions at the Observatorio del Roque de los Muchachos in
La Palma. The support of the German BMBF and MPG, the Italian INFN and
Spanish MCINN is gratefully acknowledged.
\end{acknowledgments}

\bigskip 

\end{document}